\documentclass[review]{elsarticle}
\usepackage{subfigure}
\usepackage{textcomp}
\usepackage{gensymb}
\usepackage[normalem]{ulem}

\usepackage{graphicx}
\usepackage{amssymb}
\usepackage{amsmath}
\usepackage{caption}
\usepackage{url} 
\usepackage{xcolor}
\usepackage{xr}
\usepackage{subfigure}
\usepackage{textcomp}
\usepackage{gensymb}
\usepackage{adjustbox}
\usepackage{longtable}
\usepackage{afterpage} 
\usepackage{xcolor,colortbl}
\usepackage{booktabs}
\usepackage{lineno}
\usepackage{hyperref}

\makeatletter
\newcommand*{\addFileDependency}[1]{
  \typeout{(#1)}
  \@addtofilelist{#1}
  \IfFileExists{#1}{}{\typeout{No file #1.}}
}
\makeatother

\newcommand*{\myexternaldocument}[1]{%
    \externaldocument{#1}%
    \addFileDependency{#1.tex}%
    \addFileDependency{#1.aux}%
}
\myexternaldocument{SI}

\modulolinenumbers[5]

\journal{}

\newcommand{\deltabl}{\bar{\Delta BL}}
\newcommand{\gnormunit}{mol m$^{-2}$\,\,}
\newcommand{\gunit}{J m$^{-2}$\,\,}

\begin{document}

\begin{frontmatter}

\title{A Universal Machine Learning Model for Elemental Grain Boundary Energies}
\author[UCSD]{Weike Ye}
\author[UCSD]{Hui Zheng}
\author[UCSD]{Chi Chen}

\author[UCSD]{Shyue Ping Ong \corref{cor1}}
\ead{ongsp@eng.ucsd.edu}
\cortext[cor1]{Corresponding author}

\address[UCSD]{Department of NanoEngineering, University of California San Diego, 9500 Gilman Dr, Mail Code 0448, La Jolla, CA 92093-0448, United States}

\begin{abstract}
The grain boundary (GB) energy has a profound influence on the grain growth and properties of polycrystalline metals. Here, we show that the energy of a GB, normalized by the bulk cohesive energy, can be described purely by four geometric features. By machine learning on a large computed database of 361 small $\Sigma$ ($\Sigma < 10$) GBs of more than 50 metals, we develop a model that can predict the grain boundary energies to within a mean absolute error of 0.13 \gunit. More importantly, this universal GB energy model can be extrapolated to the energies of high $\Sigma$ GBs without loss in accuracy. These results highlight the importance of capturing fundamental scaling physics and domain knowledge in the design of interpretable, extrapolatable machine learning models for materials science.
\end{abstract}


\end{frontmatter}

Grain boundaries (GBs) play an important role in determining the strength, toughness, and corrosion resistance of materials\cite{tanMicrostructureTailoringProperty2008, shimadaOptimizationGrainBoundary2002}. A key property of a GB is its energy, which determines grain growth and the GB distribution. While the GB energy can be accurately calculated using electronic structure methods such as density functional theory (DFT) calculations, the requirement for large supercells to model the inherently low symmetry GB structure limits such computationally intensive approaches to relatively small $\Sigma$ GBs. Nevertheless, substantial databases of GB energies and other properties have been developed using high-throughput DFT. For example, the GB database (GBDB)\cite{zhengGrainBoundaryProperties2020} developed by the present authors contains the calculated GB energies and work of separation of more than 50 elemental metals for both tilt and twist GBs up to $\Sigma = 9$.

Alternatively, machine learning (ML) techniques have emerged as a means to develop models that can directly predict the GB energy from compositional and structural features.\cite{kiyoharaPredictionInterfaceStructures2016,rosenbrockDiscoveringBuildingBlocks2017, snowSimpleApproachAtomic2019, gombergExtractingKnowledgeMolecular2017, echeverrirestrepoUsingArtificialNeural2014} However, existing ML models targeting elemental GBs are limited in scope by chemistry or structure type, such as face-centered cubic (fcc) Cu \cite{kiyoharaPredictionInterfaceStructures2016}, Ni \cite{rosenbrockDiscoveringBuildingBlocks2017, snowSimpleApproachAtomic2019}, Al \cite{gombergExtractingKnowledgeMolecular2017}, or body-centered cubic (bcc) Fe \cite{echeverrirestrepoUsingArtificialNeural2014} systems. These limitations are primarily a result of the choice of data source; these prior works have been developed using data sets computed using embedded atom method (EAM) potentials. While much less computationally intensive than DFT methods, EAM calculations are far less accurate, especially for non-fcc metals,\cite{zhengGrainBoundaryProperties2020} and EAM potentials are available for only a limited subset of elements. Furthermore, the majority of these prior works rely on featurization approaches such as the Smooth Overlap of Atomic Positions (SOAP){\cite{rosenbrockDiscoveringBuildingBlocks2017}, \cite{snowSimpleApproachAtomic2019}} and the pair-correlation function (PCF)\cite{gombergExtractingKnowledgeMolecular2017} that generate a large number of features which do not provide direct interpretability using commonly-used GB descriptions.

In this letter, we outline a physics-informed approach to develop a universal ML model for the GB energy of all metals, rather than for a subset of metals. We will demonstrate that the energy of small $\Sigma$ GBs of metals can be predicted to within a mean absolute error (MAE) of 0.13 \gunit using a gradient boosting regression (GBR)\cite{friedmanGreedyFunctionApproximation2001} model of the cohesive energy and four geometric GB features only. More critically, the same model can be extrapolated, without retraining, to predict the energies of high $\Sigma$ GBs with a comparable MAE of 0.12 \gunit. This work provides not only a means to rapidly predict the GB energies of any element, but also highlights the importance of choosing appropriate target normalization and features for the development of interpretable and extrapolatable ML.

The critical starting point is in re-evaluating the choice of target for our ML GB model. While prior works have attempted to directly predict the absolute GB energy, we do not believe this to be an optimal choice of target. The GB energy $E_{GB}$ is the excess energy of the GB compared to the bulk per unit area, which can be obtained from computational models as:

\begin{equation}
    E_{GB} = \frac{E_{GB,supercell} - n \cdot E^{atom}_{bulk}}{2A}
\end{equation}

where $E_{GB,supercell}$ is the energy of the supercell GB model, $n$ is the number of atoms in the GB model, $E^{atom}_{bulk}$ is the energy per atom of the bulk, $A$ is the area of the GB and the factor of 2 accounts for the fact that there are two GBs per supercell model. The physical interpretation is that $E_{GB}$ is related to the energy necessary to break or stretch bonds at the GB from their bulk equilibrium configuration. This energy to stretch or break bonds scales with the cohesive energy of the metal $E_{coh}$ (see Figure \ref{fig:ecoh}),\cite{ratanaphanGrainBoundaryEnergies2015} which ranges from $\sim 1.1$ eV atom$^{-1}$ for the alkali metals to $\sim 8.9$ eV atom$^{-1}$ for tungsten. To remove this chemical scaling effect, we have elected to use the normalized GB energy $\hat{E_{GB}} = E_{GB} / E_{coh}$ as our choice of target.

Based on the coincident-site-lattice (CSL) theory \cite{grimmerCoincidencesiteLatticesComplete1974, wolfGeometricalRelationshipTilt1989}, the GB can be specified at a macroscopic level by five degrees of freedom (DOF): two DOFs to define the plane normal of the GB (or alternatively the Miller indices $(hkl)$),  two DOFs to define the rotation axis ($[uvw]$) and one DOF to define the misorientation angle ($\theta$). Miller indices, which are defined to be integers by convention, are non-optimal for a regression task. As such, the $(hkl)$ and $[uvw]$ were converted to the inter-planar distances of the GB plane ($d_{GB}$) and the normal plane to the rotation axis ($d_{rot}$), respectively. The cosine of the misorientation angle ($\cos \theta$) was used instead of the misorientation angle itself. 

\begin{figure}
    \centering
    \includegraphics[width=1\textwidth, height=0.85\textheight, keepaspectratio]{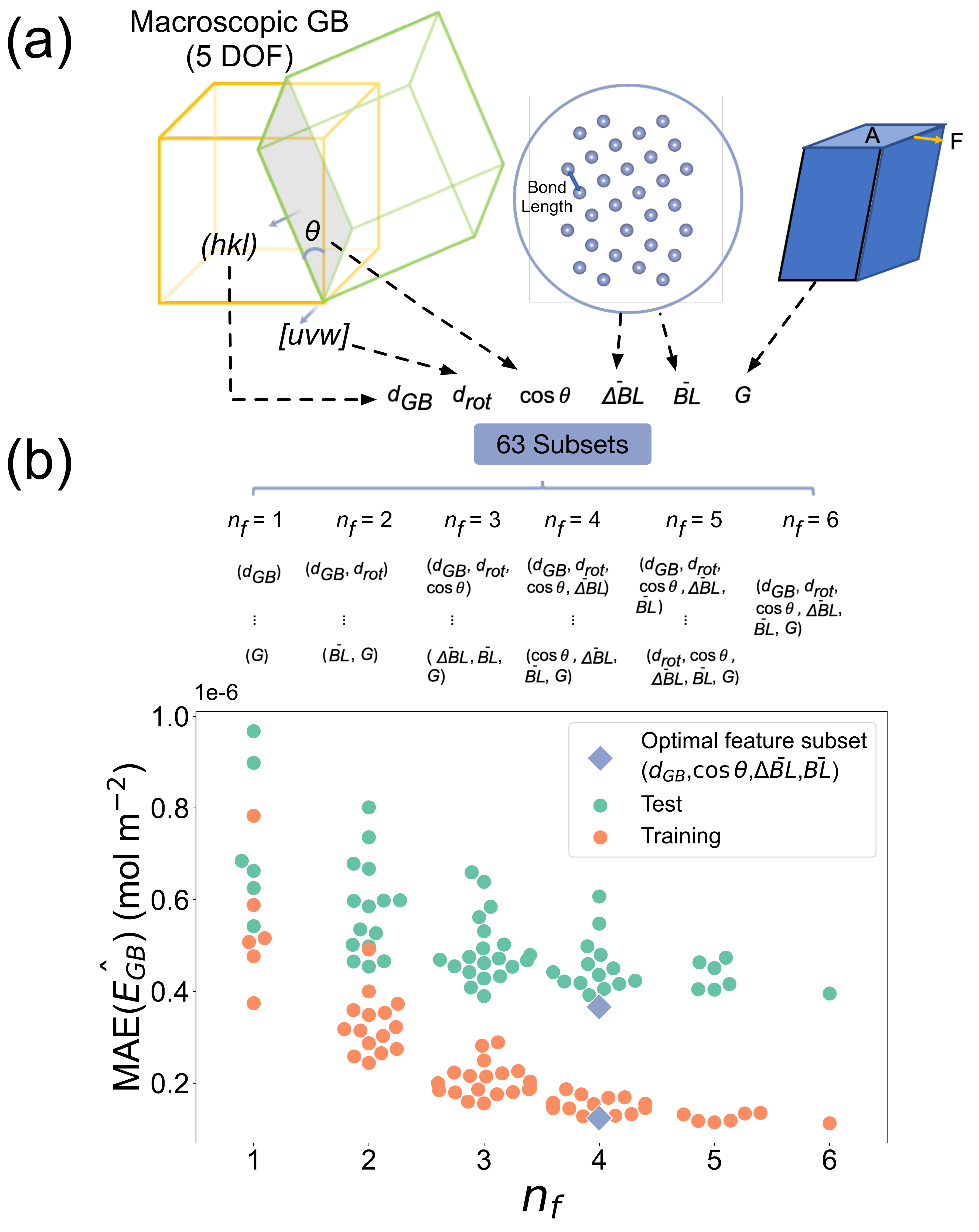}
    \caption{Feature Engineering. (a) Initial feature candidates based on the macroscopic geometry, microscopic bonding environment in the GB supercell, and the shear modulus of the elemental bulk system. (b) Best subset selection of features. There are a total of 63 possible subsets of 6 features. The swarmplot summarizes the performances of all the subsets, categorized by the number of features ($n_{f}$). The blue diamond point refers to the global optimal subset with the lowest test MAE in the predicted normalized grain boundary energy $\hat{E_{GB}}$.}
    \label{fig:feature}
    \end{figure}

In addition to these geometric GB features, we included three additional features related to bond stretching and breaking at the GB, which were partially inspired by prior works in the literature\cite{gibsonSurveyAbinitioCalculations2016}. To describe the bond deformation, we used both the average bond length in the GB supercell, $\bar{BL} = \sum_{i=1}^{n}(BL_{GB}^i)/n$, and the average change in bond lengths between the GB supercell and its bulk conventional lattice, $\bar{\Delta BL} = \sum_{i=1}^{n}(BL_{GB}^i - BL_{0})/n$, where $BL_{GB}^i$ is the bond length of the $i$th bond in the GB supercell, $BL_{0}$ is the bond length in the corresponding bulk conventional structure, and $n$ is the number of bonds counted in the GB supercell. Here, the bonds are identified by performing a local environment analysis via a Voronoi tessellation-based algorithm implemented in the Python Materials Genomics (pymatgen) package\cite{ongPythonMaterialsGenomics2013}. A positive (negative) $\deltabl$ indicates overall bond stretching (compressing) at the GB. According to the Read-Shockley dislocation model\cite{readDislocationModelsCrystal1950}, $E_{GB}$ of GBs with small misorientation angles is proportional to the shear modulus $G$. Ratanaphan et al.\cite{ratanaphanGrainBoundaryEnergies2015} have also shown previously that the GB energies of bcc Mo and Fe are related to $G$. The multi-linear regression models developed by Zheng et al. \cite{zhengGrainBoundaryProperties2020} extended this conclusion to more bcc, fcc, and hexagonal closest packed (hcp) metals. Therefore, we include the DFT Voigt-Reuss-Hill shear modulus $G$ from the Materials Project as the final feature. Figure \ref{fig:feature}(a) summarizes the preliminary set of six features considered in work. 

An initial dataset of GB energies was obtained from the GBDB\cite{zhengGrainBoundaryProperties2020}, which contains the energies of 316 GBs of 53 elements in fcc, bcc, hcp and double-hcp (dhcp) structures, after excluding Lu, Eu, and Hg due to the lack of the bulk elastic data. The $\Sigma$s of the GBs range from 3 to 9. The maximum Miller index (MMI) for the rotation axis and the grain boundary plane are 1 and 3, respectively. In this work, we extended the MMI for the grain boundary plane to 4 for 5 elements (As, Nb, Pt, Cu, and Ir), which added 5 more GBs to the initial data. Interested readers are referred to ref \citenum{zhengGrainBoundaryProperties2020} for the details on the GB structure generation and computational methods. For the model development, the 321 GBs with $\Sigma \leq 9$ were divided into training (258 GBs) and test (63 GBs) sets using stratified random sampling. The training data comprises 80\% of the GBs from each element with more than one GB, and the single GB from the remaining elements. 

A preliminary model selection process was performed in an automatic fashion utilizing a tree-based pipeline optimization tool (TPOT)\cite{olsonEvaluationTreebasedPipeline2016} with the six initial features (refer to Supplementary Materials for details). The suggested pipeline from TPOT is a decision-tree-based GBR model preceded by a polynomial feature transformation step. We recognize that there is a risk of redundancy in our selected features, e.g., $G$ has a direct relationship with $E_{coh}$, which was used to normalize the GB energy. Therefore, a comprehensive search for the ``best subset'' of the six initial features was performed using the suggested pipeline. As shown in Figure \ref{fig:feature}(b), it was found that the model's test MAE converges with only four features, ({$d_{GB}$, $\cos \theta$, $\deltabl$, $\bar{BL}$}) being the optimal set with the lowest test MAE. Further increase in the number of features leads to small increases in test MAE and decreases in training MAE, i.e., evidence of overfitting. It is worth noting that the only non-geometric feature, the shear modulus $G$, is not within the optimal subset, indicating that our proposed normalization with the cohesive energy has effectively addressed the chemical dependence of $E_{GB}$. 

\begin{figure}
    \centering
    \includegraphics[width=1\textwidth, height=0.9\textheight, keepaspectratio]{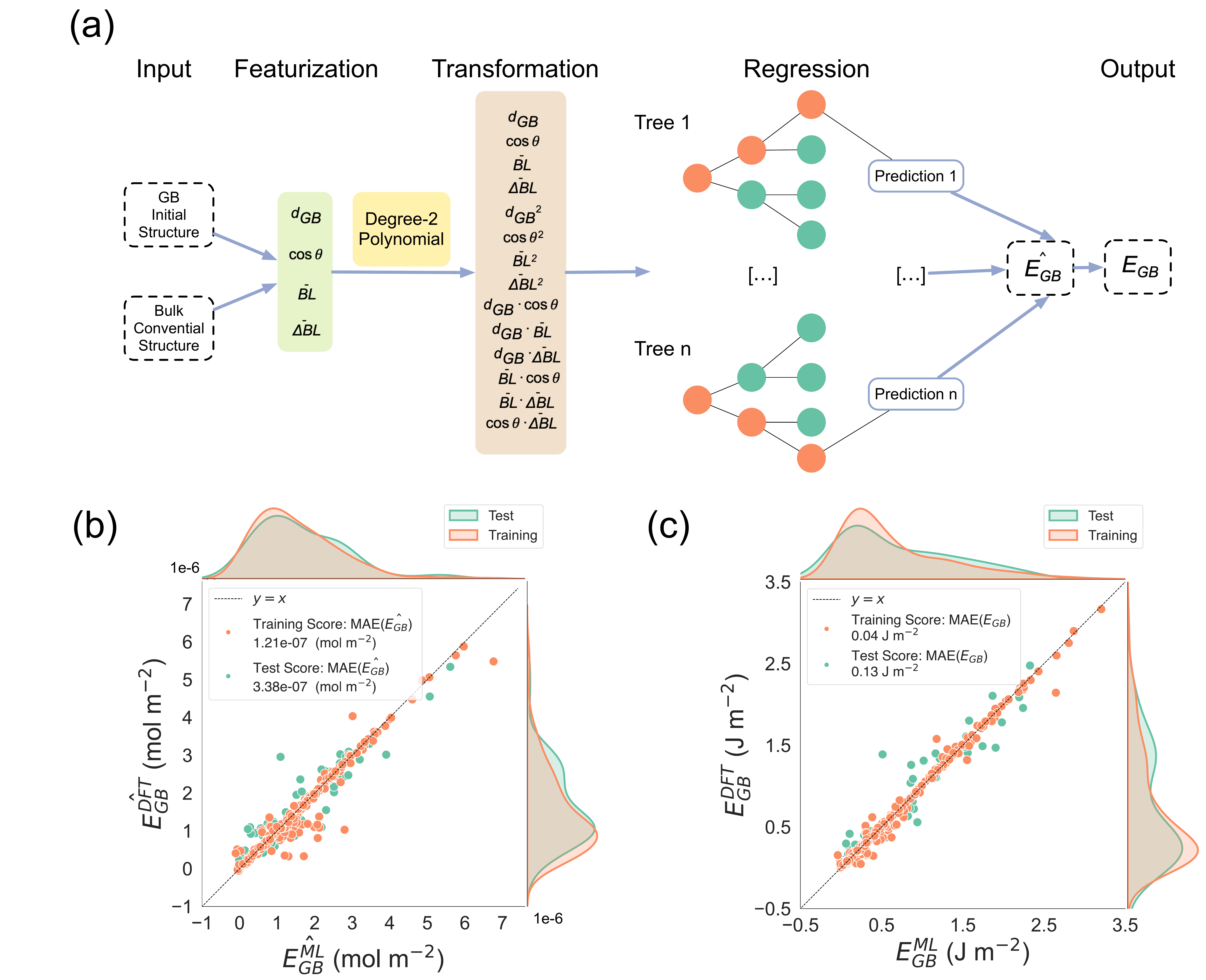}
    \caption{The machine learning pipeline and the performance. (a) The schematic illustration of the pipeline developed in this work. (b) and (c) are parity plots demonstrating the pipeline's performances on $\hat{E_{GB}}$ and $E_{GB}$, respectively.}
    \label{fig:model}
    \end{figure}

As illustrated in Figure \ref{fig:model}(a), the final optimal ML pipeline starts from the GB initial structure and the corresponding bulk conventional structure and executes the following steps: (i) featurizes the input GB structure, (ii) applies a polynomial transformation of the features up to degree 2, (iii) makes a prediction based on the trained decision-tree-based GBR model. Following this pipeline, we achieved the MAEs for $\hat{E_{GB}}$ at 1.21 $\times$ 10$^{-7}$ and 3.38 $\times$ 10$^{-7}$ \gnormunit for the training and the test data, respectively (Figure \ref{fig:model}(b)), which translate to MAEs for $E_{GB}$ at 0.04 and 0.13 \gunit (Figure \ref{fig:model}(c)). The distributions of the absolute errors for each element show that 48 out of the 55 elements have MAEs($E_{GB}$) less than 0.1 \gunit (Figure \ref{fig:error}(b)). Actinide and lanthanide elements such as Th, Ac, Yb, and Ce display the highest errors in $E_{GB}$, which may be a result of less accurate DFT calculations without proper Hubbard U values applied to account for the strong self-interaction of the f electrons. Some transition metals as Cr and Fe also exhibit higher MAEs, which we attribute to the uncertainty in the ground state magnetic ordering at the GBs in the DFT calculations.

\begin{figure}
    \centering
    \includegraphics[width=1\textwidth, height=0.9\textheight, keepaspectratio]{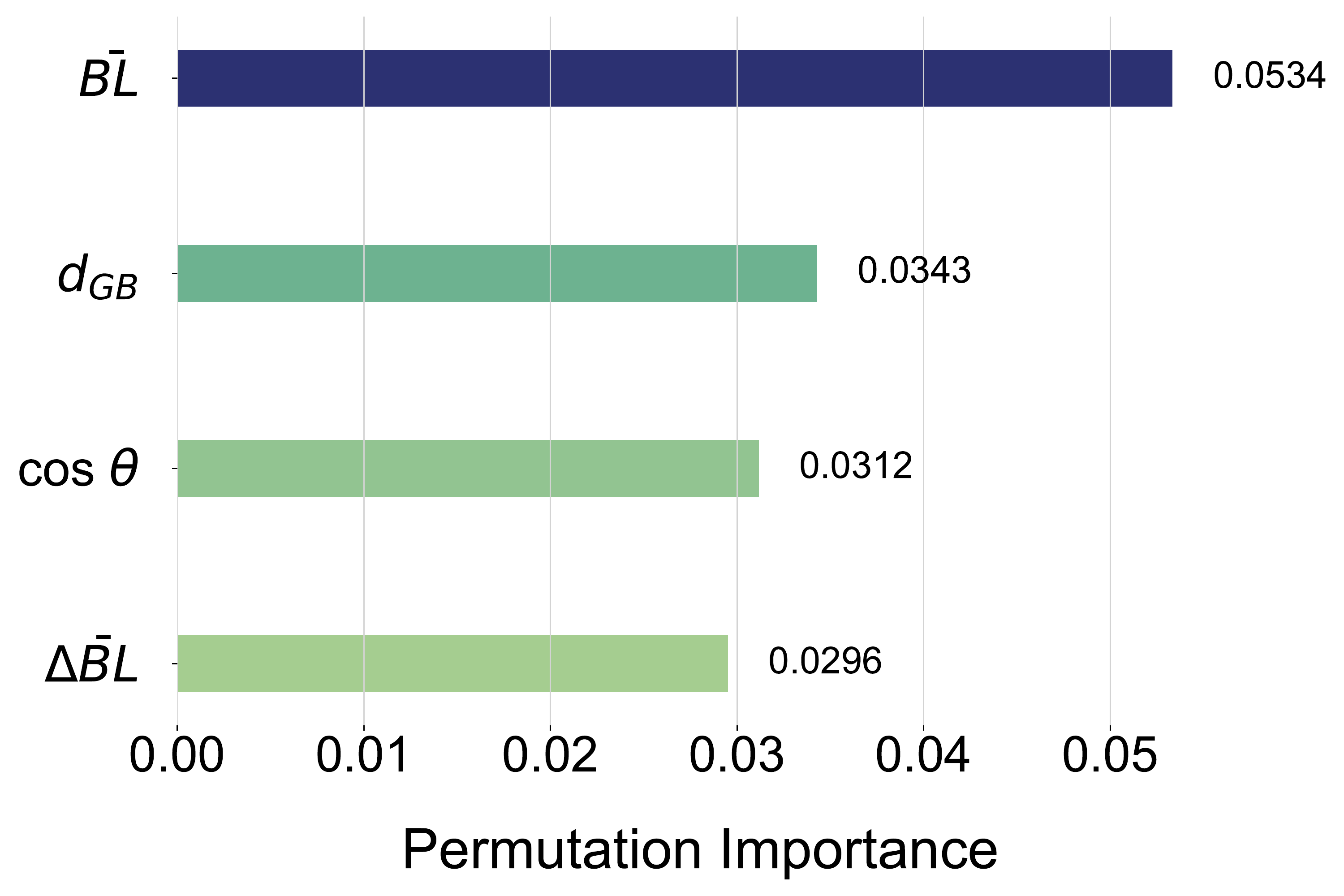}
    \caption{Permutation feature importance of the four geometric input features.}
    \label{fig:fi}
    \end{figure}

Figure \ref{fig:fi} shows the permutation importance of the four input features, which is obtained by randomly shuffling the values of one feature and calculating the decrease in the MAE of the ML pipeline.\cite{breimanRandomForests2001} The feature with highest importance is $\bar{BL}$, the average bond length, while the remaining three GB features ($d_{GB}$, $\cos \theta$ and $\deltabl$) have similar feature importances. This is consistent with previous observations by Guziewski et al.\cite{guziewskiMicroscopicMacroscopicCharacterization2021} where the grain boundary energies of silicon carbide are better described by microscopic descriptors than the macroscopic ones. Furthermore, the fact that the $d_{GB}$ has higher importance than $\cos \theta$ echos with the conclusion drawn by Rohrer et al. \cite{rohrerGrainBoundaryEnergy2011}, who found that variations in the grain boundary plane induce a greater change in the energy than the variations in the misorientation. 
 
\begin{figure}
    \centering
    \includegraphics[width=1\textwidth, height=0.8\textheight, keepaspectratio]{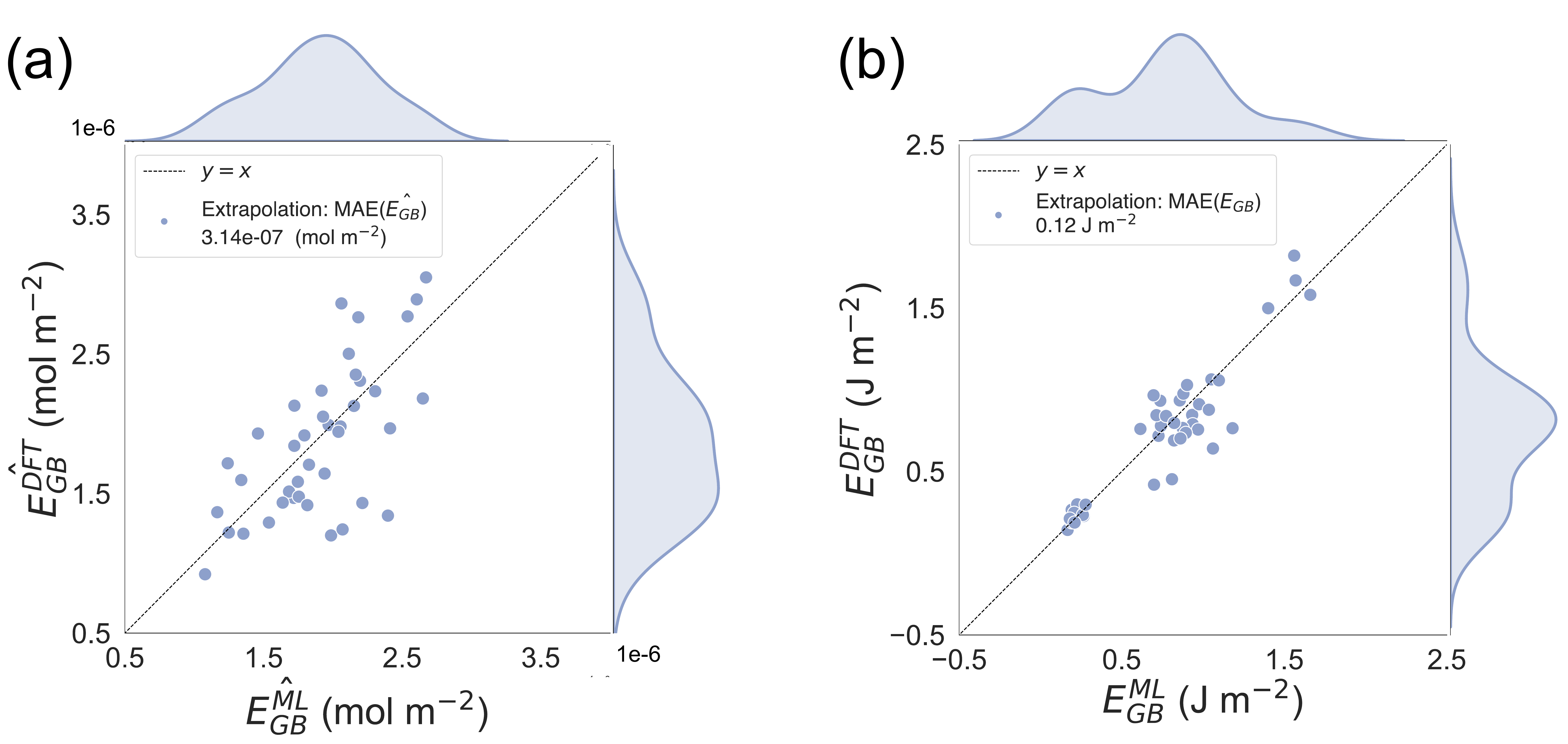}
    \caption{Performance of GB model on unseen high $\sigma$ GBs. Parity plot of predicted versus DFT (a) normalized $\hat{E_{GB}}$, (b) absolute $E_{GB}$.
    }
    \label{fig:ext}
    \end{figure}

Thus far, the optimal model presented has been trained only on low $\Sigma$ GBs that can be readily computed using standard DFT computations. If it were limited to low $\Sigma$ GBs, such a model would be of limited utility. To demonstrate its ability to extrapolate to high $\Sigma$ GBs, which require far more expensive DFT computations, an additional 40 calculations on Ta, Pd, Cu, Pt and Li GBs with $\Sigma$ from 17 to 66 with MMI of the grain boundary plane $\leq$ 8 and MMI of the rotation axis $\leq$ 1 were performed (Figure \ref{fig:data}). The model, without further retraining, achieved an MAE($E_{GB}$) of 0.12 \gunit on this data set (Figure \ref{fig:ext}(b)), commensurate with the test error of the small $\Sigma$ GB test set. 
    
To summarize, we have developed a physics-informed ML pipeline that predicts GB energies for more than 50 metals to within 0.13 J $m^{-2}$. This ML model can be applied to high $\Sigma$ GBs without loss in accuracy. A key innovation of our approach is to normalize the GB energy by the elemental cohesive energy to remove chemical scaling effects, resulting in a target that can be modelled purely using geometric and structural features. We believe this conceptual approach is general and is key to the development of ML models that are interpretable and extrapolatable. We have not attempted to model non-elemental GBs in this work given the lack of a sufficiently large dataset for model training. Nevertheless, we can speculate on the applicability of the same approach to GBs of non-elements, e.g., alloys and ceramics. Non-elemental GBs are much more complex, given that there may be compositional differences between the GB and the bulk region, e.g., preferential segregation of certain elements, etc \cite{liComplexStrengtheningMechanisms2020}. While we believe some form of target normalization with an averaged bond energy descriptor, e.g., formation energies per atom, etc., would still be useful to remove large energetic scaling effects, it is likely be less effective than in the elemental case. Furthermore, a purely geometric / structural descriptor set would not be sufficient and compositional degrees of freedom would need to be included as well. Regardless of these limitations, we believe the conceptual framework developed in this work is sound and should be extended to other properties that have a well-defined scaling relationship with bond strength, e.g., elastic constants, etc.

All GB data and models have been made available in the Github repository of the open-source Materials Machine Learning (maml) package at
\url{https://github.com/materialsvirtuallab/maml}.


\section*{Acknowledgement}

This work is supported by the Materials Project, funded by the U.S. Department of Energy, Office of Science, Office of Basic Energy Sciences, Materials Sciences and Engineering Division under Contract no. DE-AC02-05-CH11231: Materials Project program KC23MP. The authors also acknowledge computational resources provided by the National Energy Research Scientific Computing Centre (NERSC), the Triton Shared Computing Cluster (TSCC) at the University of California, San Diego, and the Extreme Science and Engineering Discovery Environment (XSEDE) supported by National Science Foundation under grant no. ACI-1053575.

\bibliography{ref}
\end{document}


\begin{frontmatter}

\title{Supplementary Information\\A Universal Machine Learning Model for Elemental Grain Boundary Energies}
\author[UCSD]{Weike Ye}
\author[UCSD]{Hui Zheng}
\author[UCSD]{Chi Chen}
\author[UCSD]{Shyue Ping Ong \corref{cor1}}
\ead{ongsp@eng.ucsd.edu}
\cortext[cor1]{Corresponding author}

\address[UCSD]{Department of NanoEngineering, University of California San Diego, 9500 Gilman Dr, Mail Code 0448, La Jolla, CA 92093-0448, United States}

\end{frontmatter}
\clearpage
\section{The scaling between $E_{GB}$ vs. $E_{coh}$}

\begin{figure}[!ht]
\centering
\includegraphics[width=1\textwidth, height=1\textheight, keepaspectratio]{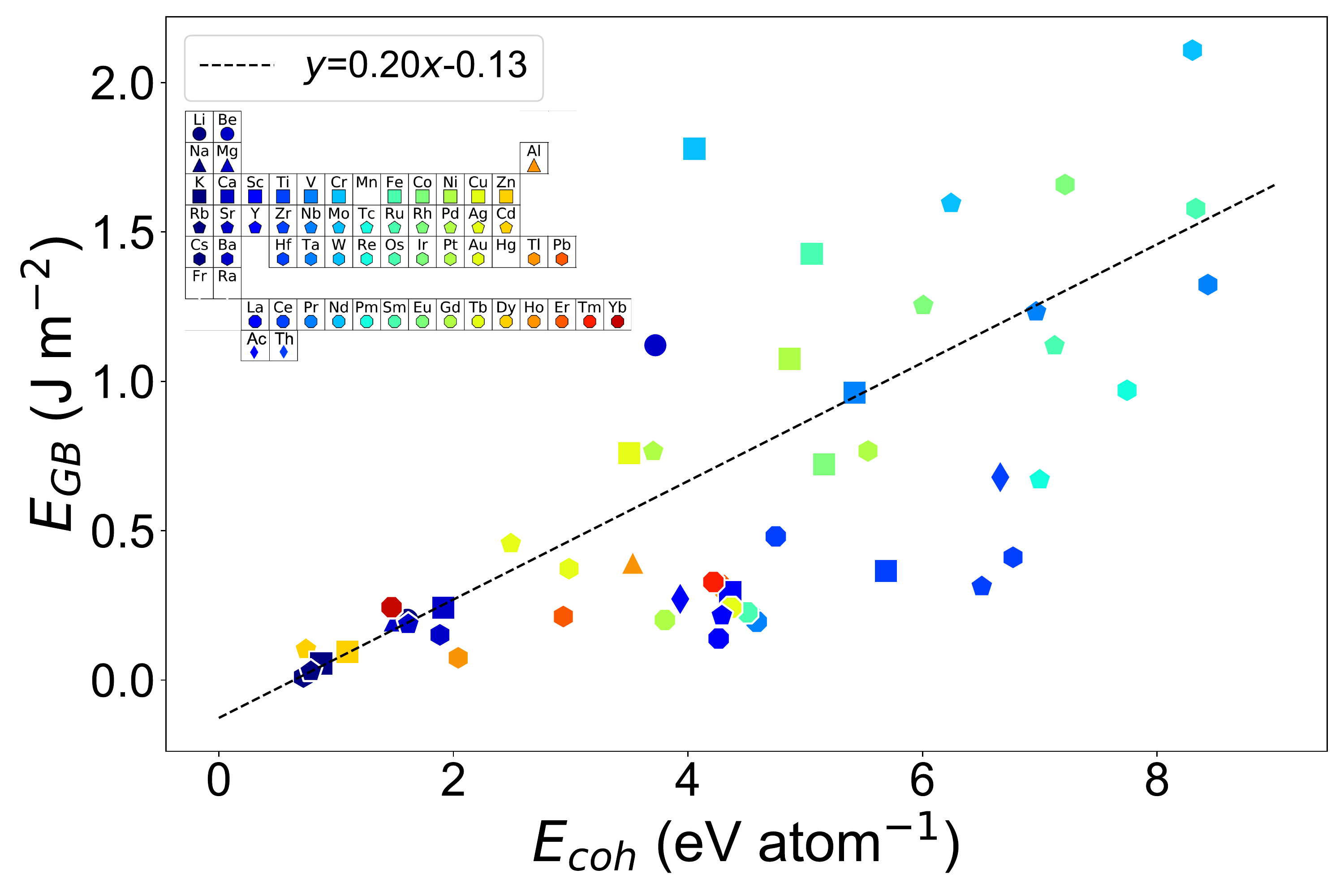}
\caption{ The averaged elemental grain boundary energy plotted against the cohesive energy. The dotted line is a fitted linear function of $y$=0.20$x$-0.13, which helps to visualize the correlation between the $E_{GB}$ and $E_{coh}$. The inset periodic table shows the marker and color scheme of the scatter plot. }
\label{fig:ecoh}
\end{figure} 

\clearpage
\section{Data Overview}

\begin{figure}[!ht]
    \centering
    \includegraphics[width=1\textwidth]{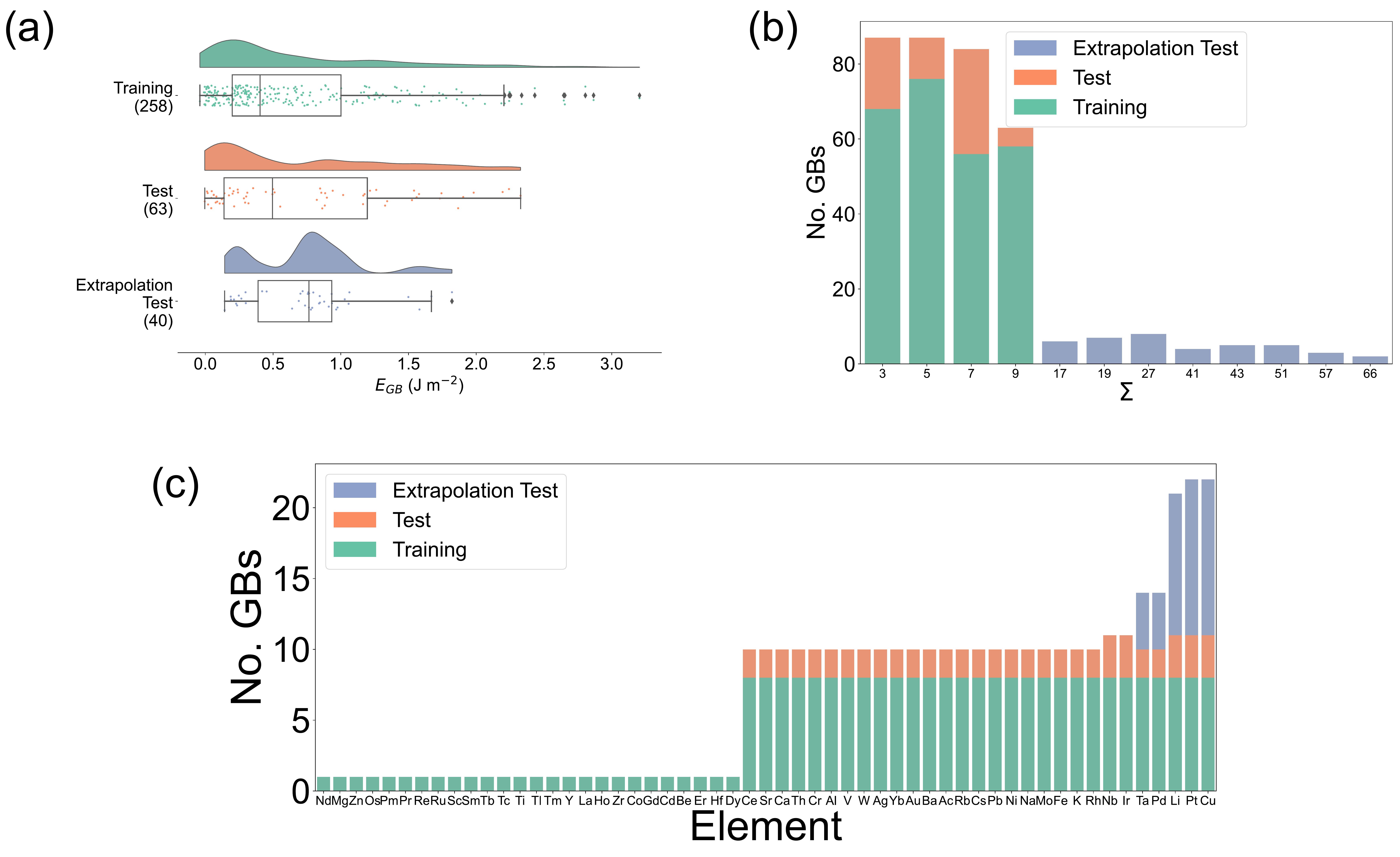}
    \caption{(a) The distribution of the $E_{GB}^{DFT}$ for different of data sets. The numbers in the bracket refer the number of data contained in the corresponding data set. (b) The $\Sigma$ distribution. For the training and test set, we only used GBs with $\Sigma \leq 9$. In addition, we also prepared an external test data, of which the $\Sigma$ ranges from 17 to 66, to test the extrapolability of the model on $\Sigma$. (c) The element distribution of the GBs.}
    \label{fig:data}
\end{figure}

\clearpage
\section{Model Development}
The optimized machine learning pipeline was selected with the aid of a tree-based pipeline optimization tool (TPOT)\cite{olsonEvaluationTreebasedPipeline2016}. Briefly, machine learning pipelines can be represented by binary expression trees with ML operators as primitives. TPOT automatically generates and optimizes the ML pipelines based on the accuracy and the complexity using genetic programming. In the current implementation of TPOT (\url{https://github.com/EpistasisLab/tpot}), the ML operators include a wide range of algorithms implemented in scikit-learn\cite{pedregosaScikitlearnMachineLearning2011} and other advanced algorithms such as decision-tree-based GBR. In this work, the population size, generations and offspring size parameters were set to 100 to evaluate a total of 10100 (100 + 100 $\times$ 100) pipelines by TPOT. Due to the relatively small data size and moderate complexity of our regression task, we confined the desired pipeline to have the structure of a regression model preceded by a data transformer (via the ``Transformer-Regressor'' template setting). Finally, the $subsample$ rate was set at 0.6 to reduce over-fitting.

The optimized model pipeline found by TPOT is an GBR model preceded by a degree-2 polynomial feature pre-processing step. The specific GBR model settings are as follows:
\begin{itemize}
  \item $alpha$ = 0.8
  \item $learning\ rate$ = 0.5
  \item $loss$ = squared error
  \item $max\ depth$ = 2
  \item $max\ features$ = 0.3
  \item $min\ samples\ leaf$ = 5
  \item $min\ samples\ split$ = 16
  \item $n\ estimators$ = 100
  \item $subsample$ = 0.95
\end{itemize}
Default values are used for all other hyper-parameters of the GBR model implemented in the open-source scikit-learn\cite{pedregosaScikitlearnMachineLearning2011} package.

\clearpage
\section{Error distribution}
\begin{figure}[!ht]
    \centering
    \includegraphics[width=1\textwidth]{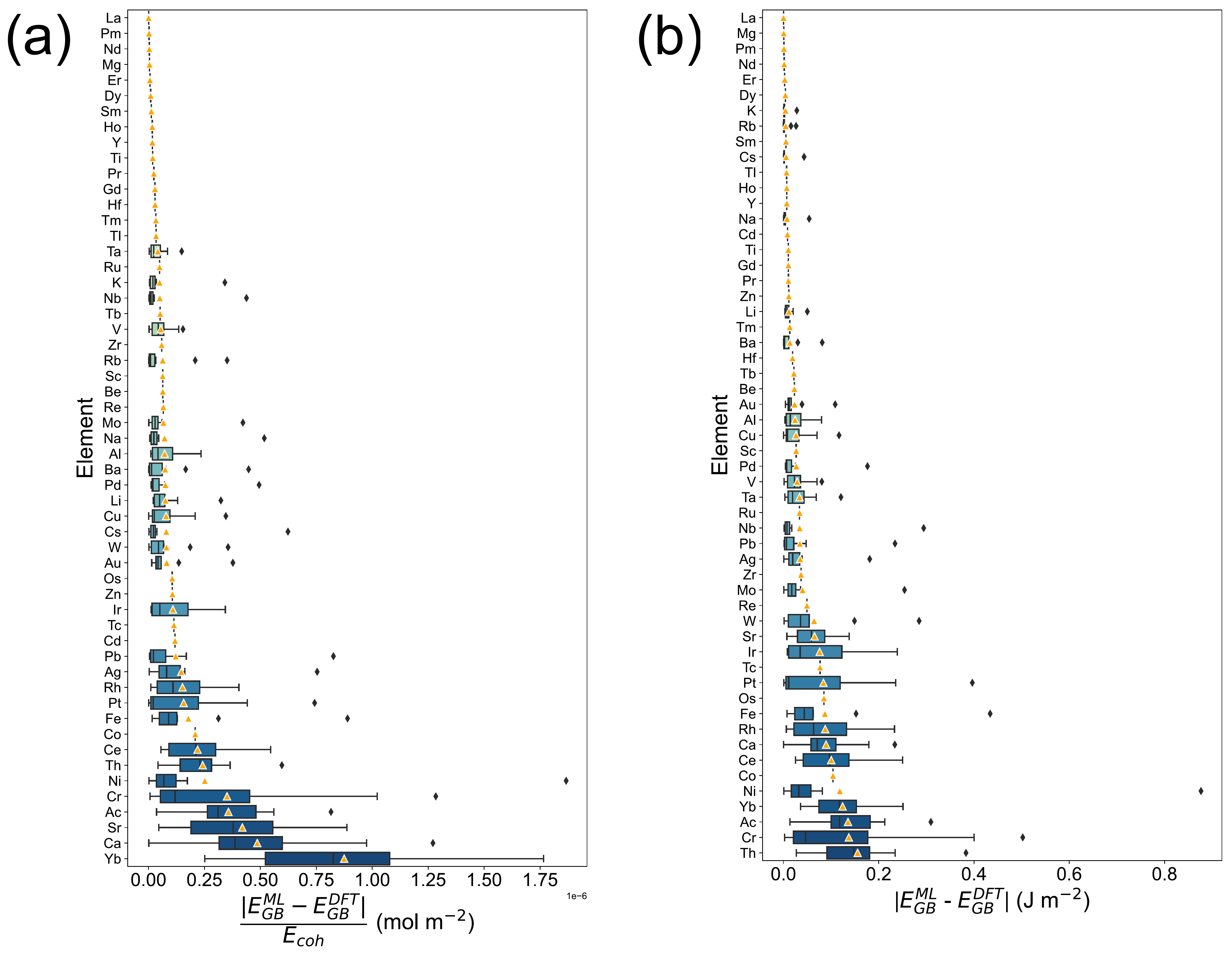}
    \caption{The boxplots showing the distributions of the absolute errors of (a) $\hat{E_{GB}}$ and (b) $E_{GB}$ for each element. The yellow triangles indicate the MAEs for each element.}
    \label{fig:error}
\end{figure}

\clearpage
\section{Feature Correlation}
\begin{figure}[!ht]
    \centering
    \includegraphics[width=1\textwidth]{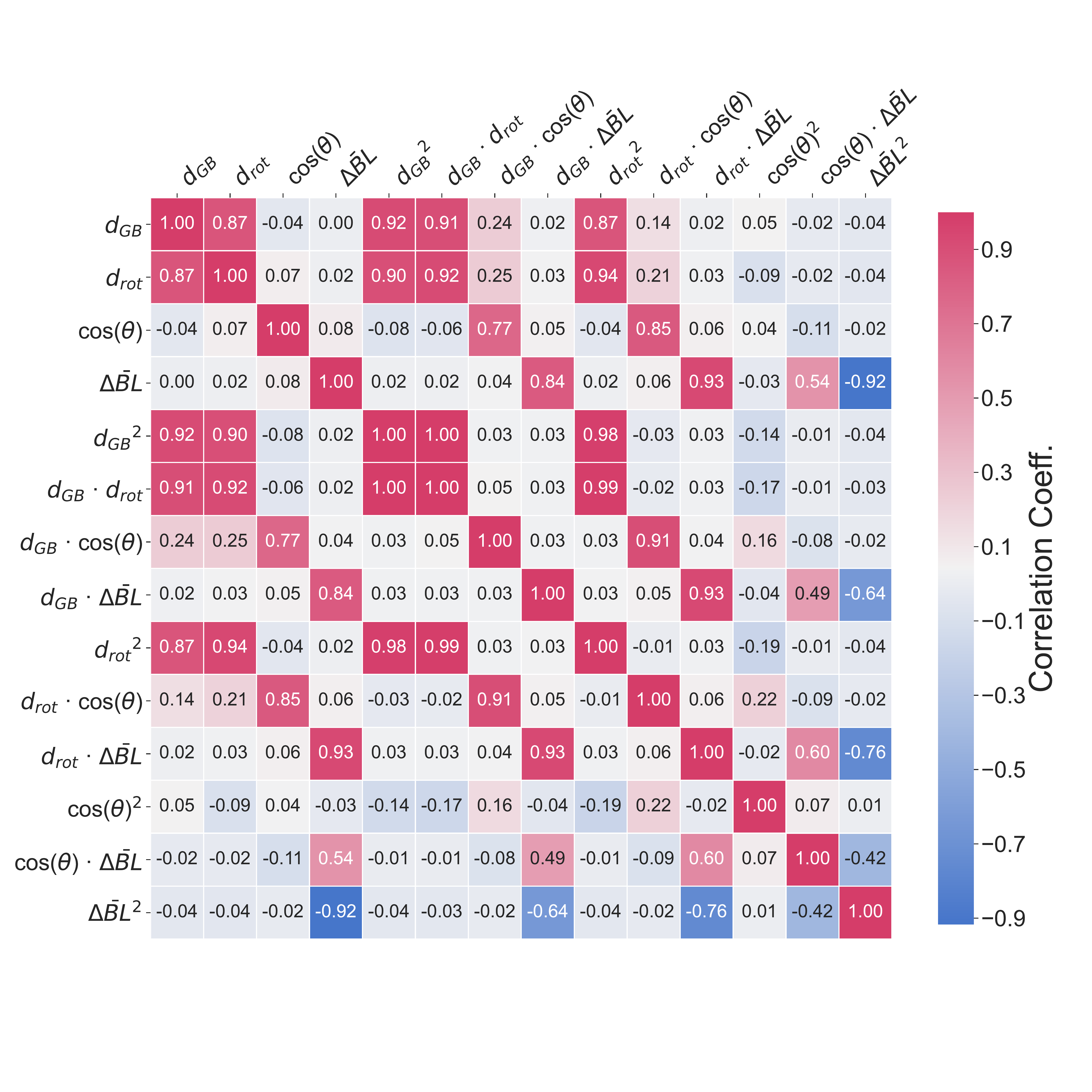}
    \caption{The Pearson correlation matrix of the 2$_{nd}$-degree polynomial terms of the optimized feature subsets. There are 17 pairs of the features that have an absolute a correlation coefficient larger than 0.75, which are considered highly correlated.}
    \label{fig:corr_14}
\end{figure}
\clearpage
\bibliography{ref}